# The lower energy consumption in cryptocurrency mining processes by SHA-256 Quantum circuit design used in hybrid computing domains

Ahmet Orun and Fatih Kurugollu

*Abstract* - **Cryptocurrency mining processes always lead to a high energy consumption at considerably high production cost, which is nearly one-third of cryptocurrency (e.g. Bitcoin) price itself. As the core of mining process is based on SHA-256 cryptographic hashing function, by using the alternative quantum computers, hybrid quantum computers or more larger quantum computing devices like quantum annealers, it would be possible to reduce the mining energy consumption with a quantum hardware's low-energy-operation characteristics. Within this work we demonstrated the use of optimised quantum mining facilities which would replace the classical SHA-256 and high energy consuming classical hardware in near future.**

*Index Terms*- CryptoCurrency, quantum computing, hash function, cryptography

**I. INTRODUCTION**

Quantum computing is one of the emerging and rapidly growing technologies in financial world [5] [20] by promising solutions for cryptocurrency mining. In particular, one of the most remarkable issues of popular cryptocurrencies [18][19] is their extremely high energy consumption during their mining process (e.g., ~10 minutes: 72,000 GW for Bitcoin). This results in spending nearly one-third of Bitcoin's current (2021) market value for energy consumption. It will also be proportionally becoming more problematic with its soaring production volume and dramatically increasing global energy shortage in the future. This critical field not surprisingly attracted several investigations and studies recently on how to make a considerable reduction of cryptocurrency mining costs [1][2][37].

Quantum computing provides fertile grounds to solve this problem as its energy consumption is extremely lower than the classical computing with CPU, GPU and ASIC [24]. As far as the quantum hardware characteristics are concerned, their size is nearly independent from their energy consumption. They consume very small energy whatever their size is in qubit capacity (as only their interface electronic instrumentation or their cooling systems consumes very low energy). There are very few previous works focused on an exploitation of quantum computing and quantum hardware utilities for such challenging hash function tasks due to Quantum Hardwares' current small size capacities [1][5]. Ablayev and Vasiliev introduces a work on quantum hashing [6] which is the core of cryptocurrency mining. Their method is based on classical-quantum where a classical bit string is used as input to produce a quantum state. Even though all scientific authorities agree on the low-energy consumption characteristic of quantum hardware, the real issue at the moment is their small-sizes and current low capacities (e.g., max 50 "reliable" qubits) to run any standard cryptocurrency algorithm like SHA-256 hashing function. The other frequently mentioned issue is probabilistic nature of quantum physics [16][17] and its adverse effect on low-level quantum hardware (*QH*) operation whereas Cryptocurrency mining process has to be very deterministic rather than probabilistic. This problem however only linked to low-level natural characteristics of quantum physics but not an issue at user level interface domain, and could be overcome by higher level supplementary methods operating in classical interface hardware (e.g. conventional dedicated PC linked to *QH*).

In this work, we investigated how any suitable "quantum hardware based" quantum SHA-256 hash function, which is equivalent to classical SHA-256, can be implemented and executed on a real publicly accessible quantum computer like IBM QX or running on a quantum simulator to demonstrate its functionality. For this aim, we implemented Quantum XOR gate (CNOT) operation in our work. Our implementation shows that SHA-256 can be utilized in QH effectively. This would lead to remarkable electricity energy saving which help to mitigate possible energy crises in future. The comparison between quantum and classical hardware energy consumptions is shown in Table 3.

The authors are with School of Computer Science and Informatics, De Montfort University, Leicester LE1 9BH, UK
Email: aorun@dmu.ac.uk, and Department of Computer Science, University of Sharjah, UAE. Email: fkurugollu@sharjah.ac.ae

## II. METHODS AND MATERIALS

*A. SHA-256 hash function*

One-way hash function or secure hash function is one of the important concepts in cryptography as used in many security related applications. This idea was firstly introduced by Merkle [7], Rabin [9] and, Diffie and Hellman [8] in their studies on public key cryptography, authentication, and digital signature. After all it became an indispensable tool for other security applications like password verification and file integrity checking as well as proof of work in Bitcoin mining.

The hash function, $H(\cdot)$, maps an arbitrary long input message, *m,* onto a fixed sized output hash code, *h (H(m) = h)* as depicted in Figure 1. The mapping function should be easily computed for any given *m*. Besides these basic properties, Merkle [7] pointed out that the following characteristics are must for a hash function and to be used in complex security applications:

- One-way or pre-image resistance: For any given hash code, *h*, it is computationally infeasible to determine the original input message, *m*.
- Second pre-image resistance or weak collusion resistance: For a given message, m, it is computationally infeasible to compute another message, *m'*, which has the same hash code, *H(m) = H(m')*.
- Strong collusion resistance: it is computationally infeasible to find a pair of messages, *m* and *m'*, such that *H(m) = H(m')*.

However, it is worth to note that strong collision is not avoidable within the generic process of hash functions but ideally kept at limited level.

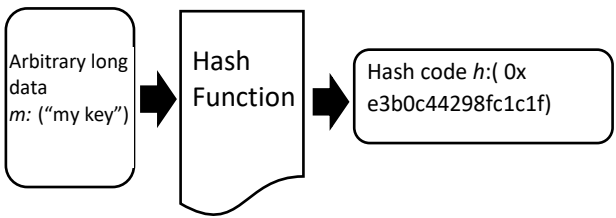

Figure 1. Basic view of a hash function as input data at varying length encrypted into a hash value at a fixed length (256 bits in the example)

After initial introduction of hash functions, National Institute of Standards and Technology (NIST) in USA developed Secure Hash Algorithm (SHA) providing security to hash functions and published as a Federal Information Processing Standard (FIPS 180) in 1993 [21]. NIST also revised SHA in 2002 and published FIPS 180-2 introducing new versions of SHA family having 256-bit, 384-bit and 512-bits hash value lengths. Among them, SHA-256 was selected as most suitable one by the cryptographic authorities with its 128 bit security level against a collision attack in comparison to SHA-512 despite its higher security level but larger performance penalty[3]. SHA-256 has also adopted for Proof-of-Work process by Bitcoin [22].

SHA-256 hash function uses the Merkle-Damgård transform based on a compression function which is applied recursively to map $n + l$ input bits to $n$ output bits [14]. For this aim, a common Davies-Meyer compression algorithm is used (Figure 2). As is a one-way compression function, its original message does not need to be restored back. A generic Davies-Meyer compression function can be defined as follows [14]:

$$f: \{0,1\}^{n+l} \to \{0,1\}^l$$
$$f(k,m) = E_k(m) \oplus m \qquad (1)$$

where *k* is the key, *m* is the message block and *E* is the encryption function with key length *n* and block length *l*. This states that the encrypted message is XORed with the message block. In SHA-256, this function is used in a recursive manner using the previous hash value, $H_{i-1}$, as the plaintext for the function to generate the next hash value, $H_i$. The message block is used as the key for *E*. Therefore the chain for the hash value in SHA-256 can be expressed as follows:

$$H_i = E_{mi}(H_{i-1}) \oplus H_{i-1} \qquad (2)$$

where; **m**: message block,
**H**: hash value (previous and next),
**E** : encryption process

At the start of process where the previous $H_{i-1}$ value is not available, the initial $H_0$ value is used as is shown in Table 1. The last H value after 64 rounds of compression function is used to determine the hash code of the message. This core process of SHA-2 is depicted in Figure 2. In each iteration of the loop, an element of 64 digit block is processed by the encryption function, *E*. As far as the specifications of the process are concerned, hash value, H, consists of 64 digit hexadecimal string. Block size of 256 bit block of message is processed in each iteration. $m_i$ shows the $i^{th}$ block of the complete message m. Here each word of string is 32 bit unsigned integer and in each iteration, a block of 64 digit message block is processed. The implementation details of SHA-256 can be found in [14]. Some examples of messages and their SHA-256 conversions are provided in Table 1.

Table I. Some of the sample messages (m) and
Their SHA-256 conversions in hexadecimal form

| Message (m) | SHA-256 conversion to 64 digit block (with 4 bit x 8x8 = 32 bit words x 8) |
|---|---|
| "This is my message for quantum" | 9b95bfa6ceb2de10d7ef3ff3b794ffea2c2ba7911a209b323a55e8f306a64931 |
| "DMU" | 3a8b4b9d4649b3573f552a9eb6b5c1244fd79815e817aa86d65422e2564b2d0a |
| "DMV" | d4fee25a1acee0e6610473456a83bd2f4f5ccf96e25c13b88f65cd79ca54d7ed |
| "A" | 559aead08264d5795d3909718cdd05abd49572e84fe55590eef31a88a08fdffd |
| $H_0$ (initial value) | 6a09e667 bb67ae85 3c6ef372 a54ff53a 510e527f 9b05688c 1f83d9ab 5be0cd19 |

The SHA-256 is characterised by XOR, bit-rotations, NOT, OR, AND, modular addition operations [14]. The core operation is XOR gate because other operations can be easily implemented by using an effective design of XOR gate. Therefore, our concern in this work is to develop and implement a Quantum equivalent of XOR gate which is CNOT quantum gate.

$a_0 |000\rangle + a_1|0001\rangle + a_2 |010\rangle + a_3| 011\rangle + a_4 |100\rangle + a_5 |101\rangle + a_6 |110\rangle + a_7 |111\rangle$

Quantum computing can be demonstrated on globally available systems like IBM QX (53 qubit as only 15 qubit accessible for public use at the moment), D-Wave (Leap) 5000 qubit, Google (53 qubit) by remote access. The facilities are optional in "*real-time* "and "*simulator*" modes. Meanwhile Microsoft provides a cloud service for Quantum programming in its own language Q# running in a simulation mode on a simulator. In terms of

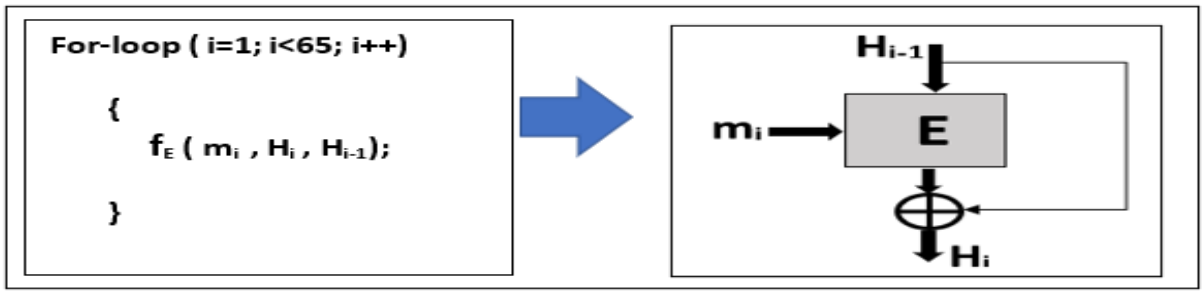

Figure 2. The continuous loop processing the chain of messages in 64-digit blocks

### B. Quantum hardware and computing

Quantum computing processes are enabled by quantum hardware (e.g., photonic, ion-trap, super conducting, topological, etc.) as called "*quantum computer*" or "*quantum annealer*" which operate according to quantum physical principles [11][12]. Quantum computing concept operationally exploits a quantum particle's two states (e.g. spin-up or spin-down, etc.) and their superposition characteristics to make calculations with an advantage of additional $2x$ states over the classical bit (0,1), as is called qubit. If qubit number is $n$, then the number of states will be $2^n$. The number of qubits refers to the capacity of quantum hardware. For example, a 3-qubit quantum computer can be described by 8-dimansional vectors like = ($a_0$, $a_1$, $a_2$, $a_3$, $a_4$, $a_5$, $a_6$, $a_7$) where each vector element, $a_i$, is a complex number coefficient. Their state descriptions would be as follows:

hardware efficiency, nowadays the most important issue of quantum hardware is high error rate of qubits as it is a current bottle neck to overcome until a new "low-error-rate" quantum hardware discovery. However, a USA company, *IonQ*, recently (October 2020) announced a new reasonable error rate quantum computer at 32 qubit capacity operates with ion-trap technique [25]. Meanwhile there is always confusion between the physical qubit and logical qubit terms, as 1 logical (low-error) qubit may refer to ~200 physical (high-error) qubits. The other disadvantageous factor is that in quantum hardware, qubits communicate to each other as an error produced by each qubit increase cumulatively over the logical gates (CNOT gate in particular) which is the main obstacle for larger quantum computers design and production. However, quantum error correction is the mainly focused area on the global basis, hence expected to be solved in near future for large scale quantum computers[26] We have to note that the quantum hardware is inevitably operated in association with the classical computers (hardware) used as an interface.

## C. CNOT quantum logic gate (XOR equivalent)

In quantum computing, CNOT logical gate (controlled-not gate) is the important one to generate quantum entanglement for the quantum tasks like teleportation, super dense coding, etc. but in our work here, it would only be used as the quantum analogue of classical computing's XOR logical gate. CNOT gate is always used by operating on two qubits at once. The first qubit operates as a "control" qubit whose state affects the second "operation" qubit's state. If the control qubit is "0" then there is no change on operation qubit's state, but if the control qubit is "1" then it changes the state of operation qubit. Control qubit state and operation qubit state are shown in Figure 3 as x-input and y-input. The classical SHA-256 can be implemented using an XOR logic gate "⊕" in its loop process which can be seen on Figure 3. Its 2-qubit quantum circuit equivalent with CNOT logic gate "⊕" where $H_{i-1} \equiv x$ and $m_i \equiv y$, as well as their logical process table is depicted at the right hand side section.

On the other hand, the SHA-256 is characterised by XOR, bit-rotations, NOT, OR, AND, modular addition operations [14]. The core operation is the XOR gate because other operations can be easily implemented by using an effective design of an XOR gate. Therefore, our concern in this work is to develop and implement a Quantum equivalent of an XOR gate which is a CNOT quantum gate.

Even though we propose a low-level quantum logical gate based operation in this study which is quantum hardware related and would be independent from any "classical" or "quantum" hash function computing approach, our proposed low-level method makes a substantial contribution to such theoretical studies to demonstrate them. The earlier studies on quantum hash function were demonstrated on both "non-binary" based [6] and "binary" quantum hashing [13]. The non-binary method accepts classical bit string as an input and convert it to quantum state output. Vasiliev introduces the technique which allows to present binary inputs by quantum states [13]. He also proved that there was reverse correlation between the characteristics of method so that, the more quantum hash function was preimage resistant, the less collision resistant it was. In their work, Ablayev and Vasiliev defined the quantum hash function by the following notifications, in which the quantum hash function $h_K$ described as [6] in Equation 3:

$$|h_K(M)\rangle = \frac{1}{\sqrt{d}} \sum_{i=1}^{d} |i\rangle \left( cos \frac{2\pi k_i M}{N} |0\rangle + sin \frac{2\pi k_i M}{N} |1\rangle \right)$$
(3)

where $M$ is the arbitrary message satisfying $M \in \{0,1\}^n$, $N = 2^n$ (for n-bit messages), $d = |K|$ and Set $K = \{k_i : k_i \in \{0,....,N-1\}\}$.

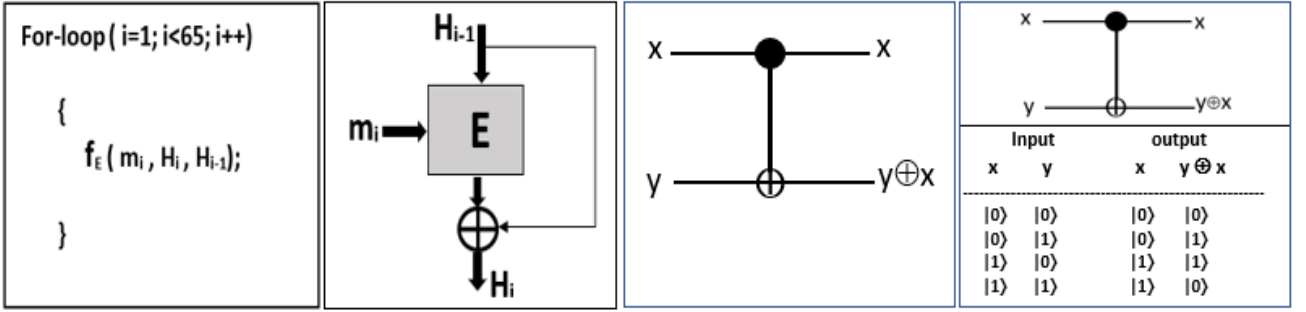

Figure 3. Implementing classical SHA-256 using Quantum CNOT gates

The further details about the quantum characteristics of the CNOT logical gate (e.g. its entanglement or superposition links, its usage with Hadamard gate, etc.[23]) are out of the scope of this work.

## III. POTENTIAL ALTERNATIVE OR SUPPLEMENTARY METHODS

### A. Quantum hash function

If δ is defined as resistance (e.g., (orthogonality) quantum collision resistance, it also corresponds to classical second pre-image resistance). For any pair of inputs W, W' (W ≠ W') it satisfies $|\langle\psi_1|\psi_2\rangle| < \delta$ where the absolute value of $\langle\psi_1|\psi_2\rangle$ inner product between two wave functions refers to distance or overlap between those two states and also refers to as fidelity. where ψ is called quantum one-way function complying with the hash function property that is easy to compute and difficult to invert. With the above notations it would be possible to evaluate formula 4 as follows [6]:

$$\left|\langle h_{M_1}|h_{M_2}\rangle\right| = \left|\frac{1}{|K|}\sum_{i=1}^{|K|}\cos\frac{2\pi k_i(M_1-M_2)}{N}\right| \leq \delta(K) < \delta \quad (4)$$

In which the pair of input massages have not to be same, $M_1 \neq M_2$.

Serial calculation of quantum hash function would be possible on globally available quantum computers or via their cloud connection. But in the latter case, logical gate calculations of a binary data string have to be made by a remote connection to paid-quantum-computing services (e.g. ~1000 shots cost about 0.25 USD) where the supplier provides a dedicated time slot for the usage. However, this would not be the ideal case since any connection between the interface domain and remote quantum hardware would have a limited security level. Hence this option has to be fortified by additional security measure for a secure connection in commercial applications.

In case of direct usage of a quantum hardware, logical gate calculations of a binary data string (e.g. input message for SHA-256 hash function) on a Quantum processor may be carried out as block-by-block of a long strings (e.g. 1000 digit long) which is normally beyond the current qubit capacity of quantum computers that are globally available (e.g. max 65 qubit). Such block-by-block string processing loops have to be run by a continuous real-time operation of quantum hardware. As this issue could be solved by partial operational division of entire task, the other quite promising global developments indicate that the quantum hardware capacity is progressively increased and targeted 1000 qubit is expected to be achieved in 2023.

*B. Quantum annealer as an alternative method for SHA-256 hash function*

Quantum annealing is linking a specific problem's best solution to a related quantum physics phenomenon to find the minimum energy state which already means the solution. To accomplish that, first, the objective function of the problem should be specified. Then calculate its energy value should be calculated by using function parameters and targeting the minimum energy point. Quantum annealing differs from the qubit gate model as the qubit gate model depends on solving the problem by controlling the logical qubit gates. The limitation of the logical qubit gate model depends on the low qubit capacity of current quantum computers globally available which is currently 64 qubit (IBM's QC at Manhattan, as Google's 72 qubit QC has some controlling issue) . Whereas quantum annealers operate depending on the evolutionary approach in which the system searches the minimum energy state corresponding to the exact solution of the objective function. In contrast to current logic gate models demonstrated by the quantum computers, quantum annealers may have huge qubit capacity (e.g. 5,760 qubits [27]) because their hardware characteristic is very different than quantum computers and limited to only optimisation problem solutions.

D-Wave$^{TM}$ for instance, produces reprogrammable quantum processing unit chips to be used in a quantum annealer. It has to work at milli-Kelvin temperature requiring a cooling system. The quantum annealer's chip is also in lattice form and demonstrates the Ising model [15] using voltages and magnetic fields to control the chip circuits. In a quantum annealer chip, the qubits are a 2D array of superconducting loops carrying electric current and behave like magnets pointing up or down, but in this case, pointing up and down at the same time according to the quantum superposition principle. To use the quantum annealer, the user maps the specific problem into a search for the lowest energy point. Then the quantum annealer processor considers all possibilities simultaneously to satisfy the qubits network of relationships with the lowest energy point. This principle would be demonstrated by minimizing the objective function (e.g. SHA-256 hash function in our case) in Figure 4 as follow:

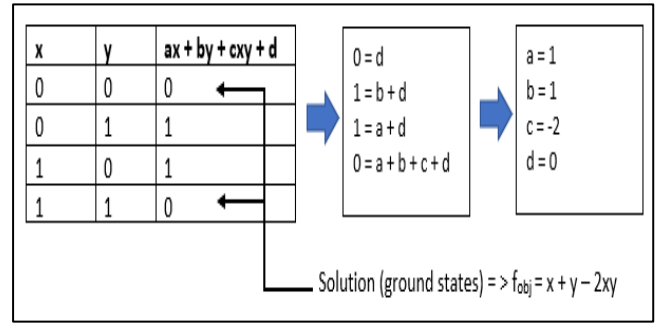

Figure 4. XOR classical logic gate or CNOT quantum logic gate logical process operated by quantum annealer with minimum energy state approach.

$$f_{obj}\ (weight,\ strength,\ qubit) = \Sigma\ weight\ .\ qubit\ +\ \Sigma\ strength\ .\ coupler \quad (5)$$

In Equation 5, the notations are described as follows:
**Coupler:** Physical device that allows one qubit to influence the others.
**Qubit:** Quantum bit participating annealing cycle and, at the end it settles down to one of the states {0,1}.
**Weight:** Qubit's tendency to jump into the final state {0 or 1}. It is constant and controlled by the user.
**Strength:** Controls the level of qubit's influence on the others. It is constant and controlled by the user.

During the quantum annealing cycle, the qubit spins keep evolving and exploring the problem space. At the end of the annealing cycle, the system reaches its ground state of the submitting problem. Then the final states are yielded as the output of the solution. A representative quantum annealing example solution for XOR or CNOT which is the core operation in SHA-256 is shown as follows:

**Problem definition:** Find minimum energy = 0 for x $\oplus$ y (XOR classical logic gate or CNOT quantum logic gate)

By the quantum annealing cycles, the system yields outputs of minimum energy states "0" which are the solution of hash function (x $\oplus$ y) as seen in Figure 4
Quantum annealers have been investigated to be used as a potential cryptocurrency process domains in future but due to current limitation of free access (e.g. $2000/hour service provided by D-Wave$^{TM}$) , they have not been used within this study.

*C. Hybrid quantum computers*

Hybrid quantum computing has been an emerging technology as a solution [29][30] to mitigate the issue of small size and low capacity of current quantum computers (QC), by merging quantum computing and classical computing hardware to utilize them in harmony. Hence, the hybrid computation may be a good choice to overcome the issues of SHA-256 capacity need in quantum computation domains. In our case the ideal approach would be, high energy consuming processes will be executed in quantum hardware and meanwhile the classical hardware will be simultaneously used in connection with QC for other tasks like: loop operations of SHA (string-by-string), the storage of the results, data format conversions (e.g. from hexadecimal to binary, etc.) and other sub-operations as seen in Figure 5.

## IV. RESULTS AND DISCUSSION

*A. Quantum circuits design to demonstrate the logical-gate level of the hash function process*

Some of the available accessible systems are shown in Table 2. The quantum circuits for the SHA-256 hash function have been designed and executed on two different domains of IBM:
-Real quantum computer hardware (Melbourne_15 Qubit)
-Quantum simulators (32qubits, 5000qubits).
as quantum simulators are specifically designed devices with different type of hardware to simulate the QC tasks. The quantum circuits exhibit only the first parts of the SHA-256 hash process loop due to the limited capacity of remotely accessible systems. The basics of circuits' NOT and CNOT gates are shown in Figure 3 and complete circuits in Figure 6 respectively.

Table II. The globally Available remotely accessible Quantum Computer Services

| System Name (Location) | Processor type | Quantum volume | Qubit capacity |
|---|---|---|---|
| Ibmq Manhattan | Hummingbird R2 | 32 | 65 |
| Ibmq Montreal | Falcon r4 | 128 | 27 |
| Ibmq Dublin | Falcon r4 | 64 | 27 |
| Ibmq Sydney | Falcon r4 | 32 | 27 |
| Ibmq Casablanca | Falcon r4H | 32 | 7 |
| Simulator stabilizer | Clifford simulator | N/A | 5000 |

Table III. Comparison between the quantum hardware and classical mining system (by different references)

| Hardware type | Energy consumption (TWh /year) | $CO_2$ Emission (Tons / year) |
|---|---|---|
| Bitcoin mining unit | 80 [31]<br>110 [32]<br>91 [33] | 267 [31] |
| Quantum computer | $25 \times 10^{-9}$ (25 kWh) [35][36] | N/A |

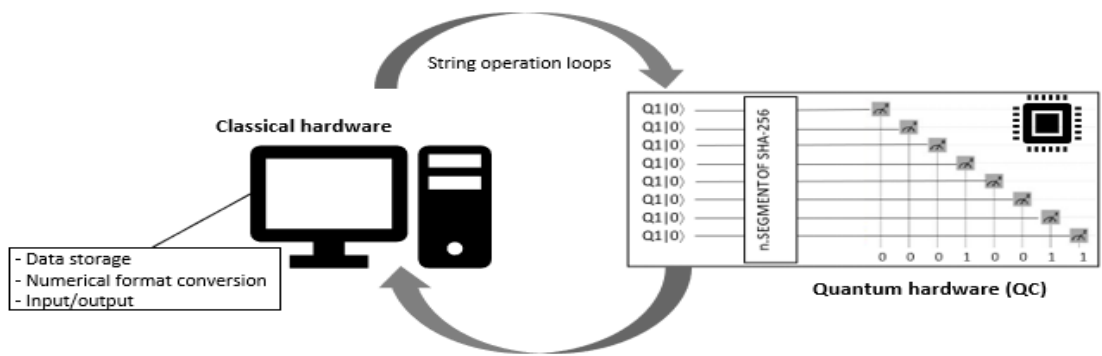

Figure 5. Proposed configuration of hybrid quantum computer design to solve the issues of current QCs low capacity by sharing SHA-256 tasks categorised as "high-energy consumption (in QC)" and "low energy consumption (in Classical hardware)"

It has been reported that there are about 1 million Bitcoin miners globally available [34]. The related calculations of electricity consumption and $CO_2$ gas emission for both (quantum and classical systems) can be made as shown in Table 3.

Practically SHA-256 hash function quantum "qiskit" codes can be directly executed on the real quantum computer hardware remotely without the need for circuit design as shown in Figures 6 and 7.

(measurements) include 16 x"0" corresponding to $H_0$ ($h_0$) and $m_1$ section (top of the circuit) where no measurements are made. In Figure 6, the histogram at the top-right exhibits 1022 shots of the logical gate operation measurements for 24 qubits (0-23) whose result is calculated for qubits 23-16 as "0,1,1,1,1,1,1,1" and "0s" for the remaining qubits 15-0 means that no any logical gate operation is made for this range. As already known, the quantum computer's calculation style is based on probability, hence the results are yielded after number of shots (loop) which

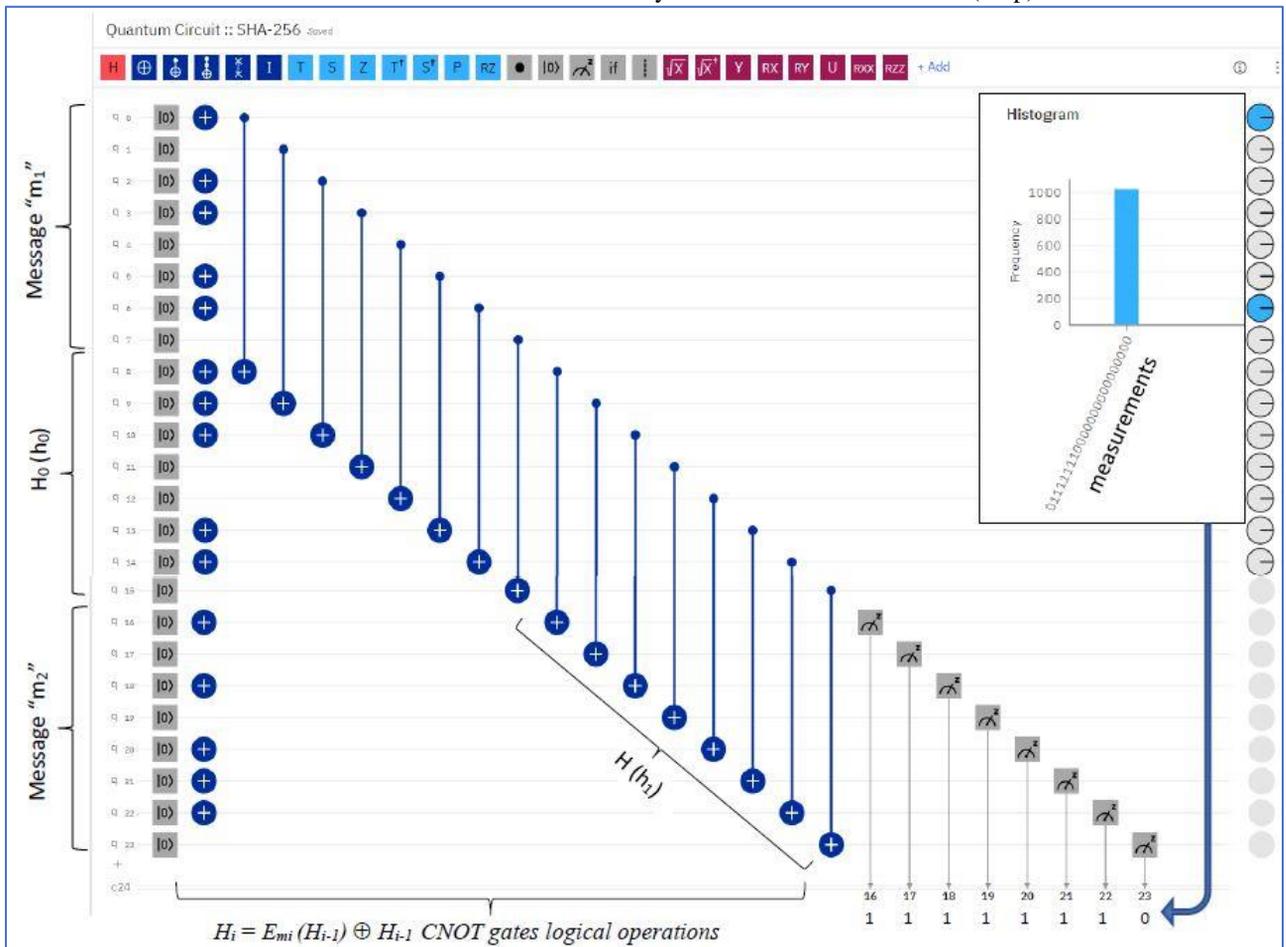

Figure 6. SHA-256 Circuit design which demonstrates for-loop (first 8-bit) of SHA-256 hash function in 5000-qubit capacity Quantum Simulator. Its measurement results are also shown at top-right window.

The circuit processes initial (starting from the end) part of message m= "this is my message for quantum" and initial $H_0$ value as;

$H_0(h_0)$= 0110101000010011110011001100111 which is equivalent of h0= 0x6a09e667 in hexadecimal form is shown in Figure 6. We have to note that in the process only $H_0(h_0)$ is taken instead of $H_0(h_7)$ which is normally aligned with the end of message "m". The results the results are yielded after number of shots (loop) which corresponds to number of repetitions of the quantum circuit. In the circuit of quantum simulator (Figures 6), the NOT gate is shown as "⊕" which is also known as Pauli-X gate. It changes $|0\rangle$ state to $|1\rangle$ state. As is shown at the left edge of the circuit, all qubits initially start with $|0\rangle$ states for values of "0", and then the values of "1" are generated by use of "NOT" gates which refer to $|1\rangle$ states. In this way we can make any data entry of any binary string into the quantum circuit. In the middle region of circuit we use CNOT gates (Controlled-NOT gate corresponds to classical XOR logic gate) which link between two binary values to process XOR logical operation between them. The measurement icons at the bottom-right corner of the circuit just display the output of calculations. The measurement result "Message"m2" is yielded after the CNOT logical gate operation between "Message"m1" and "$H_0(h_0)$" as described in Formula 2. The circuit of real quantum computer as shown in Figure 7 (other than the quantum simulator) contains the same components with the Figure 6, but with different numerical configuration. The "most probable" result is displayed by the frequency

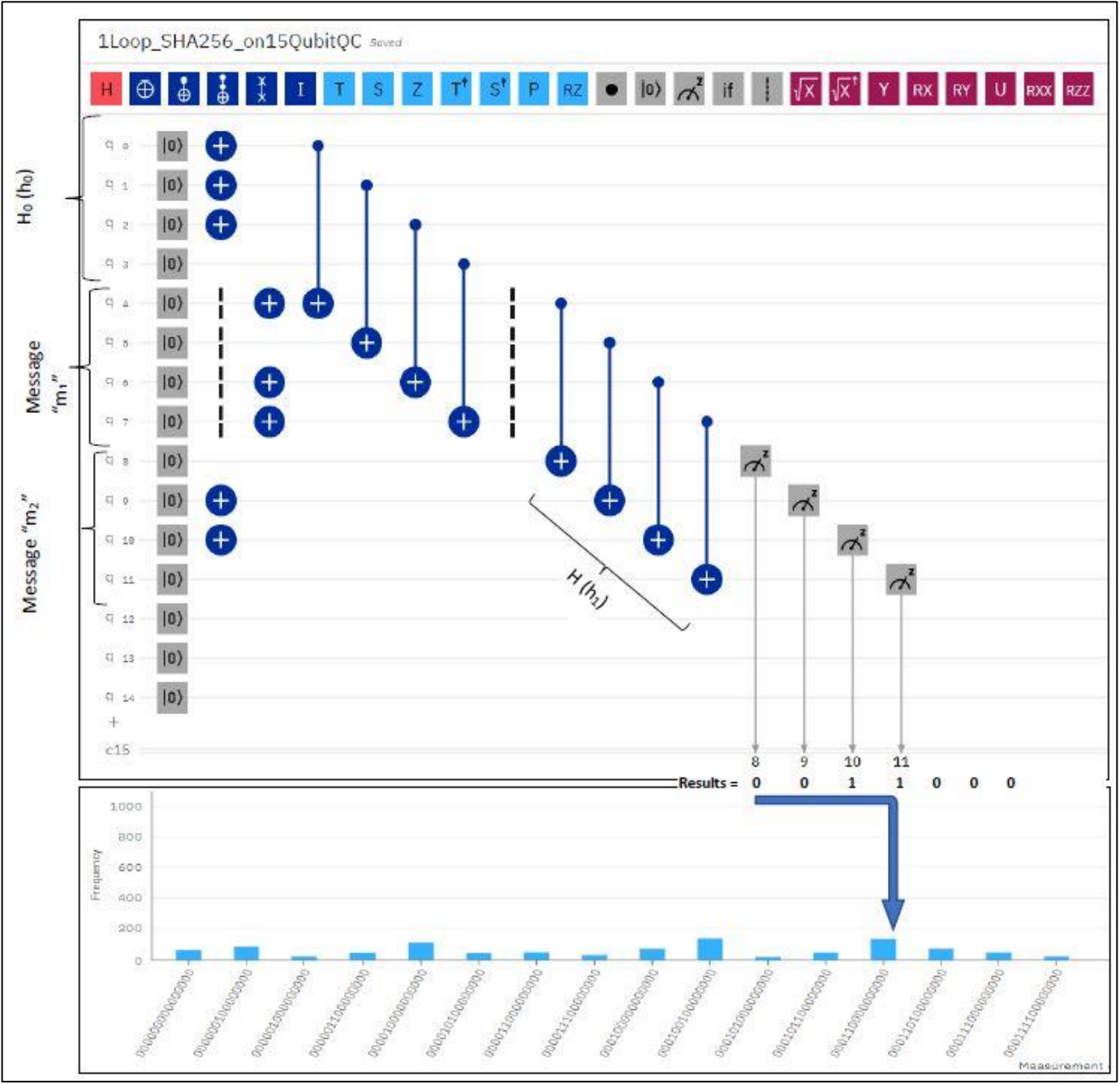

Figure 7. (**Top**) Single loop of SHA-256 hash function operation by the circuit where i=4bit (as explained in Figure 3) is executed on IBM's 15qubit real quantum computer hardware (Melbourne). (**Bottom**) The results are quantum probabilistic and different than simulator ones. The most probable results are shown by blue arrow in the frequency histogram table whose probability values are generated after 1022 shots of quantum units. histogram (bottom side) corresponding to "Message"m2" is "0,0,0,1,1,0,0" for the qubit range 14-8 respectively. The remaining "0" values for the qubit range 7-0 means that no any logical gate operation is made for this range.

## V. CONCLUSION

Within this study, the basics of the classical SHA-256 hash function which is the most energy consuming part of cryptocurrency mining have been demonstrated at a proof-of-concept level. Even thought the whole mining process requires a very high volume of quantum gate combinations, it fortunately has the identical repetitive structures and configurations whose functions to be proven by a small size of globally accessible quantum computers. As far as the energy consumption of the proposed system is concerned, the interface hardware for QCs and/or classical computing components of a hybrid computers have not accountable energy consumption since they are only utilized for data format conversions, data storage, quantum logical process loop operations, etc. The other popular concern is about quantum physic's probabilistic nature and is reflected by quantum hardware's high error rate (CNOT gate in particular) which has also been proven as is not a problem at a higher level of host system's operation where the probabilistic results are further re-interpreted by a classical (interface) hardware to make the results deterministic enough for Cryptocurrency mining process. It has to be noted that the results obtained in Figure 7 rely on high-error rate quantum hardware, whereas there is currently lower error rate hardware like IonQ's 32-qubit quantum hardware ($QH$) utility and IBM's 65-qubit ($QH$) utilities available but at some service cost. However more reliable and larger quantum hardware would be soon presented in near future (e.g. 1000 qubit QC is targeted by 2023 and 1M qubit QC is expected by 2030) this is due to high level of globally challenging competition between the quantum institutions and private companies.

On the other hand, as there would be a common debate in financial community based on a speculative idea claiming that the easier production of Cryptocurrency may lead to its sharp price fall and lead to corruption of whole crypto currency system. Whereas some finance authorities agree that the cryptocurrency prices do not rely on any absolute factor but rather manipulated only by global supply-demand chain. In addition to this, we have to note that in our study in fact we are not proposing any quantum algorithm to speed up the current mining process nor make it easier. Our proposed method is just to make the current mining process less energy consuming which helps reduce global carbon emission, avoiding a global waste of electricity, initiating further developments of quantum finance area, etc.